\documentstyle[12pt]{article}
\newcommand{\ga}{\gamma}
\newcommand{\si}{\sigma}
\newcommand{\beq}{\begin{equation}}
\newcommand{\eeq}{\end{equation}}
\newcommand{\lsim}{\raise.3ex\hbox{$<$\kern-.75em\lower1ex\hbox{$\sim$}}}
\newcommand{\gsim}{\raise.3ex\hbox{$>$\kern-.75em\lower1ex\hbox{$\sim$}}}

\author{{\sc   Ju.~M.~Khalack$^{(1)}$, V.M.~Loktev$^{(1)}$, 
A.B.~Nadtochii$^{(2)}$,} \\ 
{\sc I.V.~Ostrovskii$^{(2)}$, 
H.-G.~Walter$^{(3)}$ }\\[0.5ex]
\it $^{(1)}$ Bogolyubov Institute for Theoretical Physics,\\
\it National Academy of Sciences of Ukraine,\\
\it 14-b Metrologichna Str., Kyiv 252143, Ukraine
\\[0.5ex] 
\it $^{(2)}$ T.~Shevchenko Kyiv University \\
\it 6 Glushkova Ave., Kyiv 252127, Ukraine \\[0.5ex]
\it $^{(3)}$ Friedrich Schiller University, \\
\it Institute for Optic and Quantum Electronics,\\ 
\it 07743 Jena, Germany}
\title{NONLINEAR DISLOCATION DYNAMICS 
AND CRYSTAL SONOLUMINESCENCE}

\begin{document}
\sloppy
\setcounter{page}{1}
\maketitle

\begin{abstract}

The sonoluminescence of ionic semiconductors were studied. 
The main attention is paid to threshold phenomena which accompany the 
light irradiation, namely --- point defect creation and nonlinear 
ultrasound wave attenuation. 
The model for description of processes under investigation which 
connects the sonoluminescence excitation with the onset of point 
defects (vacancies and intersticials) generation by moving under 
ultrasound action screw dislocation with a jog. 
The attempt is made to estimate the parameters of crystals which 
define the jog motion in its cristal relief.

\end{abstract}

\sloppy
\section{Introduction}

The problems of dislocation
dynamics, especially nonlinear one, 
and defects in crystals are
still of high interest 
\cite{Sudzuki,ChCh96,GI:FMM,Sugakov,Natsik}.
Among them there is  well-known 
effect of acoustic waves on ionic crystals 
(mainly, semiconductors) which is studied 
and investigated in details (see, for example, 
\cite{MethUS,BGGKR91,KTP93,P93}). 
Sound and ultrasound (US) treatment 
results in change of 
the various important characteristics of 
semiconducting media, which, in its turn, 
can depend upon amplitude of acoustic waves. 
The most interesting here are those 
phenomena, when the changes induced by 
such waves have threshold character, 
i.e., are observed when wave amplitude reaches 
the certain value.  One of 
these brightly threshold phenomena is 
sonoluminescence (SL), which was discovered 
by Ostrovskii et al.\ \cite{Disc} 
(see also \cite{Ostr:mon}) and
represents a glow of ionic crystals, 
subjected to an US load of an overthreshold amplitude. 

The analysis of SL spectra has allowed to 
establish \cite{Ostr:mon}
that the large role in SL excitation is played 
by the crystal point defects of crystal, 
number of which essentially increases  
above the threshold. 
In such a situation it was natural to suppose generation of point 
defects, which can be stipulated by 
the motion of dislocations, to be the reason for threshold.
As a whole the sequence of processes can be such that, that US 
shakes dislocations (edge and screw)  available in a crystal, 
the amplitude of their motion being 
proportional to the amplitude of US wave.  Thus, free segments 
of the dislocations between the pinning points  are  oscillating only. 
However any new defects cannot be generated by them.  

There are several 
ways known for point defects to be created \cite{Friedel}, 
one of which is climbing of jogs on screw dislocations. Other 
ways (for example, 
intersection with dislocations of the ``forest'') in conditions of 
rather small density of the dislocations should be less effective. 

In the present paper an attempt is made to consider 
the threshold phenomena 
connected just to generation of defects by a driven jog on a screw 
dislocation. An 
equation of such a motion is investigated for US amplitude 
value up to a threshold and after it. 
Experimental study on amplitude relations of US damping 
in ionic crystals is also conducted.  
The results obtained are compared to the theory.

\section{Model of Nonlinear Dynamics of 
Dislocation with a Jog}

\subsection{Approach and Equations}

The jogs on a screw dislocation can be regarded as some kind 
of pinning points, which, however, in difference from usual 
ones, can at some conditions 
move together with its ``own'' dislocation. 
This motion is not free, and any 
displacement of the jog between its initial position and nearest, 
final, one is always 
accompanied by a creation of a point defect --- vacancy or 
interstitial. Already this 
implies that such positions appear nonequivalent, and 
consequently the potential $W_{\rm jog}(y_{\rm jog})$  
of a jog proposed 
in Ref.~\cite{prepr:pd} in difference, 
for example, with a 
potential  of Peierls relief is not symmetrical under 
translations (see Figure 1).  Just 
this potential determines the motion of the jogs in a crystal, 
which we shall consider below. 
We should notice only that the energy of the point defect creation 
makes usually a few 
eV; so the thermal overcoming of appropriate barriers 
by a jog is 
improbable. Therefore here the essential role should be played 
by forces due to 
oscillating dislocation segments,  or, in other words, by forces 
of linear tension. 

Let us consider the simplest case of a segment of a screw 
dislocation of 
length $2L$ with a jog in the middle, the ends of which are fixed 
on so-called strong 
(i.e., immovable) pinning points.  Then (in a neglect by a Peierls 
relief that is fair for the relatively 
high temperatures) the motion of the free segments of a dislocation 
in US field is 
described by well-known model of an elastic string \cite{Granato}; 
corresponding equation of a motion  
may be written in the form: 
\beq
M_{\rm dis}\frac{\partial^{2}y}{\partial t^{2}} +
B\frac{\partial y}{\partial t} -
T_{\rm dis}\frac{\partial^{2}y}{\partial x^{2}} 
= -\si_{zx}^{'}(t)\,b,
\label{o1}
\eeq
where $x$ is a coordinate along a dislocation, 
$y (x)$ is transversal displacement, $t$ is 
the time, 
$M_{\rm dis}$ is a dislocation mass per unit length, 
$B_{\rm dis}$ is a factor of a friction, 
$T_{\rm dis}$ is a linear tension of a dislocation string, 
$\bf b$ is a Burgers vector, 
$\si_{zx}^{'}(t)$ is appropriate (active) component of 
stress deviator, caused by US. 

Let the jog be placed in a point $x=x_{\rm jog}$. 
A condition of a dislocation pinning 
down (at small US amplitudes) is then 
\beq
y(x_{\rm jog})=0
\label{o2}
\eeq
At large amplitudes of an external force (i.e., of 
the exciting US wave) the condition (2) 
should be replaced by an equation of a jog motion, 
which can be written as 
\beq
M_{\rm jog}
\frac{\partial^{2}y_{\rm jog}}{\partial t^{2}} +
B_{\rm jog}
\frac{\partial y_{\rm jog}}{\partial t} +
\frac{\partial W_{\rm jog}}{\partial y_{\rm jog}}
=T_{\rm dis}
\left(
\left.\frac{\partial y}{\partial x}\right|_{x_{\rm jog}+0} -
\left.\frac{\partial y}{\partial x}\right|_{x_{\rm jog}-0} 
\right),
\label{o3}
\eeq
where $M_{\rm jog}$ and $B_{\rm jog}$ 
are a jog mass and friction coefficient for it, 
correspondingly (they, as well as $M_{\rm dis}$, $B_{\rm dis}$, 
are phenomenological parameters), and 
$W_{\rm jog}(y_{\rm jog})$ is Loktev-Khalack potential (see Figure 1). 
In general case its form depends upon all previous positions of a jog.
For example, if the last has moved from a position A on Figure~1 to 
position B with a vacancy being created, the branch 
AB$^{'}$C$^{'}$D$^{'}$ has ceased to exist. In other words, the jog 
can return ``initial geometry'' only passing exactly to C$^{'}$ 
on the curve $W_{\rm jog}(y_{\rm jog})$, for what it needs to 
overcome a potential barrier, appropriate to the interstitial 
formation.  
Thus, it is essential that the energies of the formation 
of the vacancies and interstitials are definitely 
different; it naturally makes 
a potential relief 
$W_{\rm jog}(y_{\rm jog})$  central-asymmetrical one and 
is reflected in the jog motion under the action of US wave.

\subsection{Threshold characteristics}

Let an US wave with the amplitude of acoustic displacement 
$u_{\rm ac}$ and frequency 
$\omega_{\rm ac}$ be spreaded in a crystal. 
Then the force per unit length exerted on the 
dislocation (see (1)) is
\beq
-\si_{zx}^{'}(t)b=
f_{\rm or}\frac{\omega_{\rm ac}u_{\rm ac} b }{v_{\rm us}}
\cos{\omega_{\rm ac}t}=\si_{\rm us}b
\cos{\omega_{\rm ac}t},
\label{o4}
\eeq
where $f_{\rm or}$ 
is some factor, dependent on the orientation of an US wave (on its 
polarization and the direction of propagation in a crystal), 
and $v_{\rm us}$ is a sound 
velocity. 
Under the action of force (4) free segments of a dislocation
begin bowing out and in 
accordance with the increase of $u_{\rm ac}$ a situation will set in, 
when the amplitude of 
this bowing out (and consequently, pulling forces of a linear 
tension exerting on 
the jog) will become sufficient for the jog to overcome the 
potential relief.  The 
amplitude of this force is determined (see a right term in Eq.(3)) 
by condition 
\beq
\left|\frac{\partial W_{\rm jog}}{\partial y_{\rm jog}}
\right|_{\rm max}
=T_{\rm dis}
\left(
\left.\frac{\partial y}{\partial x}\right|_{+0} -
\left.\frac{\partial y}{\partial x}\right|_{-0} 
\right)_{\rm max}. 
\label{o5}
\eeq

The solution of equation (1) together with (5) 
for the case of external force (4) 
gives the following expression for the threshold 
amplitude of US produced displacement: 
\beq
u_{\rm ac}^{\rm thr}(\omega_{\rm ac})= 
\frac{v_{\rm us}}{\omega_{\rm ac}}
\frac{M_{\rm dis}L}{8 f_{\rm or}b T_{\rm dis}}
\left[
I_{1}^{2}(\omega_{\rm ac})+I_2^2(\omega_{\rm ac})
\right]^{-1/2}
\left|
\frac{\partial W_{\rm jog}}{\partial y_{\rm jog}}
\right|_{\rm max},
\label{o6}
\eeq
with the substitutions 
\beq
I_1(\omega_{\rm ac})=
\sum_{n=0}^{L/2b}
\frac{
\Omega^{2}_{n} -\omega_{\rm ac}^{2}
}{
\left(
\Omega^{2}_{n} -\omega_{\rm ac}^{2}
\right)^{2}
+\left(
\omega_{\rm ac}\Gamma_{\rm dis}
\right)^{2}
},
\label{o7}
\eeq
\beq
I_2(\omega_{\rm ac})=
\sum_{n=0}^{L/2b}
\frac{
\omega_{\rm ac}\Gamma_{\rm dis}
}{
\left(
\Omega^{2}_{n} -\omega_{\rm ac}^{2}
\right)^{2}
+\left(
\omega_{\rm ac}\Gamma_{\rm dis}
\right)^{2}
},
\label{o8}
\eeq
and 
\beq
\Gamma_{\rm dis}=B_{\rm dis}/M_{\rm dis},
\label{o9}
\eeq
where
\beq
\Omega^{2}_{n} = 
\frac{\pi^{2}(2n+1)^{2}T_{\rm dis}
}{M_{\rm dis}L^{2}}
\label{o10}
\eeq
are the eigen-frequencies of the dislocation segments (n=0,1,...). 
The obtained expressions (7), (8) testify  
that at $\Gamma_{\rm dis} < \Omega_0$ the 
threshold characteristics defined by (6), 
should have resonant character (see Figure~2).

\subsection{Point defect generation}

The asymmetrical form of a curve 
$W_{\rm jog}(y_{\rm jog})$ stipulates that basic conclusion, 
that a threshold condition (5) for the jog motion  with 
a formation of vacancies begins to hold 
still before an appropriate condition for the motion with 
interstitials formation. Then it 
is easy to see that for the amplitude $u_{\rm ac}$ between these two 
threshold values the 
jog can move and move during an only one halfperiod of 
every period of the external force action. 
It means that in the given range of US amplitudes the jog 
drift happens only in one direction, and a creation of one vacancy 
corresponds to each its climb.  

Already from physical reasons 
it is clear, that the appearance of 
vacancy gives premises for lattice relaxation 
in a vicinity of a defect, which with necessity 
should be accompanied by an acoustic emission.  
It is important that such an emission happens before transition of 
a jog to an oscillatory mode.   
On the other hand, transfer of the jog into some 
new position at corresponding value $u_{\rm ac}$ causes a reduction 
of resultant forces of a linear tension along the $x$ axis.  
As a result 
this total force will decrease so that further jog climbing 
will become impossible, and the jog will stay in new, biased rather 
initial, equilibrium position. 

However, such a displacement has also a positive effect, 
if a reduction of threshold 
US amplitude necessary for interstitial formation is 
to be mentioned. This 
threshold value is determined by the same formula (6), 
in which instead of value 
of derivative $\partial W_{\rm jog}/\partial y_{\rm jog} |_{\rm max}$, 
appropriate to interstitial formation, stands a 
half-sum of appropriate derivatives for two opposite 
directions; 
the observable threshold for full oscillatory jog motion is given by  
\beq
u_{\rm ac}^{\rm thr}(\omega_{\rm ac})= 
\frac{v_{\rm us}}{8\omega_{\rm ac}} \, 
\frac{M_{\rm dis}L}{ f_{\rm or}b T_{\rm dis}} \,
\displaystyle
\frac{
\left|
\frac{\partial W_{\rm jog}}{\partial y_{\rm jog}}
\right|_{\rm max}^{i}+
\left|
\frac{\partial W_{\rm jog}}{\partial y_{\rm jog}}
\right|_{\rm max}^{v}}{
\sqrt{
I_{1}^{2}(\omega_{\rm ac})+I_2^2(\omega_{\rm ac})}}.
\label{o11}
\eeq

Thus, after a beginning of large-amplitude oscillatory 
jog motion (i.e., after 
overcoming by the amplitude $u_{\rm ac}$ of its 
threshold value (11)) continuous 
generation of defects of both types begins. 
The total nunber of defects generated by jog per  
one US wave period above the threshold 
grows proportionally to  
US amplitude.

\subsection{US attenuation}

Because of real crystals containing the dislocations of different 
length, let us consider a pure single-crystal with a network of 
dislocations, as well as a certain quantity of point defects, being 
the weak pinning centers for the former ones.  It is assumed also 
that initial concentration of the point defects is small enough for 
the mean distance $L_{c}$ between them to be of the order of the 
network length $L_{N}$: $L_{c}^{(0)}\gsim L_{N}$. 

If the acoustic stress of small amplitude is applied to
a crystal, the dislocations are bowing out between pinning
points, which results in amplitude-independent US
attenuation. At the higher stresses a breakaway occurs,
giving rise to hysteresis losses, and consequently to
the increase of US attenuation (see \cite{Granato}). 

But in the  case under consideration ($L_{c}^{(0)}\gsim L_{N}$)
the increase of attenuation due to dislocations 
unpinning from 
the weak pinning centers may be negligible, so that
the amplitude dependence of the US attenuation is
determined by the motion of the jogs on the skrew dislocations. 
When the amplitude of an acoustic stress is sufficient for a
creation of the vacancies, the jog merely changes its
equilibrium position, not causing hysteresis losses by
itself. 
The attenuation is changed by the newly created
vacancies. 
The last serve as the additional weak pinning centers for
the dislocations, so that the attenuation coefficient for low
frequency range is given by Granato-L\"{u}cke expression
\cite{Granato} 
\beq
\alpha_{\rm H}
(\omega_{\rm ac}, \si_{\rm us})
 = 
\frac{\omega_{\rm ac}}{2 v_{\rm us}}
\frac{\Lambda L_{N}^3}{L_{c}}
\frac{8\mu b^2}{\pi^4 T_{\rm dis}}
\left[
\frac{\pi f_{m}}{4b \si_{\rm us}L_{c}}-1
\right]
\exp{\left(-\frac{\pi f_{m}}{4b \si_{\rm us}L_{c}}\right)}, 
\label{012}
\eeq
where $\Lambda$ is the dislocation density, $\mu$ is a
shear modulus, $f_{m}$ is the maximum value of the
binding force. 
Here the value of $L_{c}$ is no longer equal to the
initial value $L_{c}^{(0)}$, but depends on the amplitude
of acoustic stress $\sigma_{\rm us}$. 
In the assumption that the vacancies are uniformly
distributed throughout the volume, $L_{c}\approx
(N_{v}(\sigma_{\rm us})+(L_{N})^{-3})^{-1/3}$. 
The number of vacancies per unit volume is 
\beq
N_{v}(\sigma_{\rm us}) = 
\int\limits_{L^{v}(\sigma_{\rm us})}^{L_{\rm max}} 
N_{\rm jog}(L)
\frac{Y(L,\sigma_{\rm us})}{y_0} d L,
\label{o13}
\eeq
where $N_{\rm jog}(L)$ is the distribution function for
the dislocations with jogs, $Y(L,\sigma_{\rm us})$ is the
displacement of a jog from its initial position in the US
field, $y_0$ is a lattice spacing, and $L^{v}(\sigma_{\rm
us})$ is the minimum half-length of a dislocation, the jog
on which can climb at given value of $\sigma_{\rm us}$. 
If we suggest that $N_{\rm jog}(L)=N_{\rm jog}={\rm
const}$, then 
\beq
N_{v}(\sigma_{\rm us}) = 
\frac{2 N_{\rm jog}L_{\rm max}^{3}}{3\pi^2 T_{\rm dis}}
\frac{(\sigma_{\rm us}-\sigma_{\rm us}^{v})^2 
(2\sigma_{\rm us}+\sigma_{\rm us}^{v})}{\sigma_{\rm us}^2}.
\label{o14}
\eeq

For stresses high enough for the jog transition into 
an oscillatory mode (i.e., for 
$\sigma_{\rm us}>\sigma_{\rm us}^{\rm thr}$) the
attenuation of US is determined by the losses due to
creation of point defects. 
The attenuation coefficient for this case is  
\beq
\alpha_{\rm jog}(\si_{\rm us})
=
\frac{\omega_{\rm ac}}{2 v_{\rm us}}
\mu 
\left(W_{v}+W_{i}\right)
\int\limits_{L^{thr}(\sigma_{\rm us})}^{L_{\rm max}} 
N_{\rm jog}(L)
\frac{\Delta y_{\rm jog}(L,\sigma_{\rm us})}{y_0
\si_{\rm us}^2} d L, 
\label{015}
\eeq
where $W_{v}$ and $W_{i}$ are correspondingly the energies 
of creation of vacancy and interstitial, 
$\Delta y_{\rm jog}(L,\sigma_{\rm us})$ is the amplitude
of the jog oscillations, and $L^{thr}(\sigma_{\rm
us})$ is the minimum half-length of dislocation, the jog
on which is oscillating. 
If we adopt the above assumption  $N_{\rm jog}(L)={\rm
const}$, the amplitude dependence of attenuation is given
by the factor
\beq
\frac{(\sigma_{\rm us}-\sigma_{\rm us}^{thr})^2 
(2\sigma_{\rm us}+\sigma_{\rm us}^{thr})}{\sigma_{\rm us}^4}
\label{o16}
\eeq
(the account is taken of proportionality of 
$\Delta y_{\rm jog}(L,\sigma_{\rm us})$ to the difference
$\sigma_{\rm us}-\sigma_{\rm us}^{thr}$). 

The further increase of acoustical stress can activate the
Frank-Read sources, as well as additional slip planes (in
accordance with the factor $f_{\rm or}$ in (4)), the last
giving rise to new threshold-like pecularities in the
amplitude dependence of US attenuation. 

The behaviour of the attenuation during the unloading
cycle is determined by the newly created defects. 
Additional dislocations lead to the increase of losses,
while the point defects reduce losses to some extent. 
It is noteworth that the amplitude dependence of US
attenuation below the threshold in the case of unloading
is given by the expression (12) with the constant value
$L_{c}$, because the jogs are fixed at some new positions
and cannot create any point defect.
The qualitative form of the curve of the amplitude dependence of US 
attenuation during the loading and unloading cycles 
is shown at Figure~3.

\section{Experiment and Analysis}

The monocrystallyne samples NaCl, KCl, and ZnS
(sphalerite) were experimentally studied at room
temperature. 
The initial density of dislocations in crystals under
investigatin not subjected to US treatment made for NaCl
about $10^4$cm$^{-2}$ and for ZnS --- about $5\cdot
10^3$cm$^{-2}$. 
Longitudinal US waves were excited by piezoceramic
transducers of PZT type within a frequency range
$1.5\,$MHz$\,\lsim \,(\omega_{\rm ac}/2\pi)\, 
\lsim \,7\,$MHz. 
The factor $\alpha$ of US attenuation was measured by
two techniques: 1)~by comparing of the exciting US
rf-voltage $V$ with those picked up from receiving
transducer, and 2)~with the help of a probing pulse by an
echo-pulsing method. 
The US waves were excited by a continuous rf-voltage. 

Typical dependence of US attenuation $\alpha$ on US
amplitude taken from the sample KCl-1A is
shown on Figure~4. 
Along X axes a rf-voltage $V$ of the frequency
$f=(\omega_{\rm ac}/2\pi)=2.5$MHz is given. 
US amplitude is proportional to this voltage. 
It is seen that at a low amplitude $V$ 
the attenuation of US wave
is equal to some value (point A) which remains practically 
constant at $V$ increase up to a point B ($V\approx 10$
V). After this value of $V$ the attenuation begins to
decrease (part BC of the curve). 
This reduction can be explained by the threshold vacancy
generation as it was described in Section~2.4. 

Part CD of this curve corresponds to attenuation growth
when the generation of vacancies as well as interstitials
begins (cm.\ parts BC and CD of the theoretical curve at
Figure~3). 
It is interesting to notice that according to the
calculations the very onset of vacancies generation is
accompanied by some small attenuation increase. 
It can be interpreted as a result of competition between
hysteresis losses grows due to the dislocations breakaway
near the ``old'' pinning centers and generation of new
ones (vacancies), which cause free dislocation segments
slowing doun. 
It seems that observable peak at the part AB of the
experimental curve can be qualitatively ascribed to the
effect reminded. 
The relatively strong additional growth of attenuation
(part DE) is possibly provoked by activation of
dislocations with jogs motion in another slip plane, what
results in one more threshold value \cite{Kh:UFZh}. 

Figure~5 represents the results of the same study of the
sample KCl-2 under US of $f=1.75$~MHZ. 
There were two cycles of a loading. 
In the first case a voltage raised up to a point C below
the threshold (curve~1), and then lowered to zero
(curve~2). 
In the second case $V$ increased up to a value 
above the threshold (point D$^{'}$, curve~3).
It is seen that for both samples a hysteresis character
of attenuation dependencies is evidently observed. 
However the concrete shape of the curves depends directly
on the maximum amplitude of the acoustical wave 
propagating in the sample. 
If the last is larger than the threshold one, then the
US attenuation during the unloading is greater than
that for the loading because of additional
dislocation density. 
But if the maximum US amplitude is less than the
threshold one (though is sufficient for vacancy 
generation),
then the unloading curve is found below the loading one,
in accordance with the greater number of weak pinning
centers present at the moment. 

As a whole, observable US attenuation meets rather good
qualitative agreement with the predictions of the model
proposed. 
One of the most important ones among them is the
coincidence of the threshold of the SL excitation with
that of the point defect generation. 
Figure~6 comfirms this supposition: such an equality, in
fact, takes place (see curve~1 for SL and curve~2 for US
attenuation, both of which are sharply changed at the
same value of US amplitude). 

As to quantitative agreement of experimental and
theoretical curves, it should be emphasised that their
shape depends strongly on the dilocation length
distribution, the character of which is not known for
samples under investigation. 
The single quantity to be derived from this studies
exactly is a ratio of the threshold values of US amplitude
for continuous point defects generation and for creation
of vacancies. It gives a ratio of activation energies of
interstitial and vacancy creation by a jog  
by means of the following expression, 
as it easily can be obtained from (6) and (11):
\beq
\frac{W_{i}^{*}}{W_{v}^{*}}=
2\frac{u_{\rm ac}^{\rm thr}}{u_{\rm ac}^{v}}-1.
\eeq

Results of our experiments shown on the Figure~4 and~5 give
this ratio to be about 3. 
Indeed, as it seen from these figures, $(u_{\rm ac}^{\rm
thr}/u_{\rm ac}^{v}) \sim 2$, where $u_{\rm ac}^{\rm thr}$
corresponds to 1, and $u_{\rm ac}^{v}$ is $\approx 0.5$
in relative units. 
The correlation between $W_{i}^{*}$ and $W_{i}^{*}$ well
satisfy the values found from radiation experiments 
\cite{Luschik}.

\section{Conclusions}

The main results of the given work can be formulated as
follows:

1.~The model for point defect generation by moving under
US wave action jog on skrew dislocation is proposed. 
It predicts the threshold character of this motion which
is defined by energies of creation of the vacancy or
interstitial. 
It must be however noticed that in the simplest above
approximation we restricted ourself to one sort of
vacancies and one sort of interstitials. 
In fact, the jogs in real crystals can generate different
point defects of one type which can differ, for example,
by their charges (valencies). 
Namely that can be a reason for observation of various
impurity center spectra as SL.

2.~The US wave nonlinear absorption as a function of US
amplitude reveals before threshold minimum which was
explained above by supposition that new (generated by
jog) point defects (vacancies) become the additional
pinning centers for free dislocation segments. 

3.~The experimental study of some 
semiconducting compounds showed that observable
dependencies neet satisfactory agreement with predictions
of the model developed. 
It allows to estimate some energetical crystal parameters
which are in agreement with data obtained from
independent investigations. 

\section*{Acknowledgement}
We kindly acknowledge the Bundesministerium f\"ur Bildung, 
Wissenschaft, Forschung und Technologie, Deutchland, and the ISSEP 
Program of International Soros Foundation for partial financial 
supporting of the project.

\newpage

\section*{Figure captions}


Figure~1: Potential relief of jog in a crystal. 
The right and left sides of this relief correspond to jog
motion with generation of vacancies and interstitials. 

Figure~2: The dependence of threshold US amplitudes upon
its frequency. 
The upper and lower curves describe the threshold values
for interstitials and vacancy generation,
correspondingly.

Figure~3: The general behaviour of US absorption under
loading and unloading produced by US wave. 

Figure~4: Experimental behaviour of US absorption by 1st
KCl sample. 

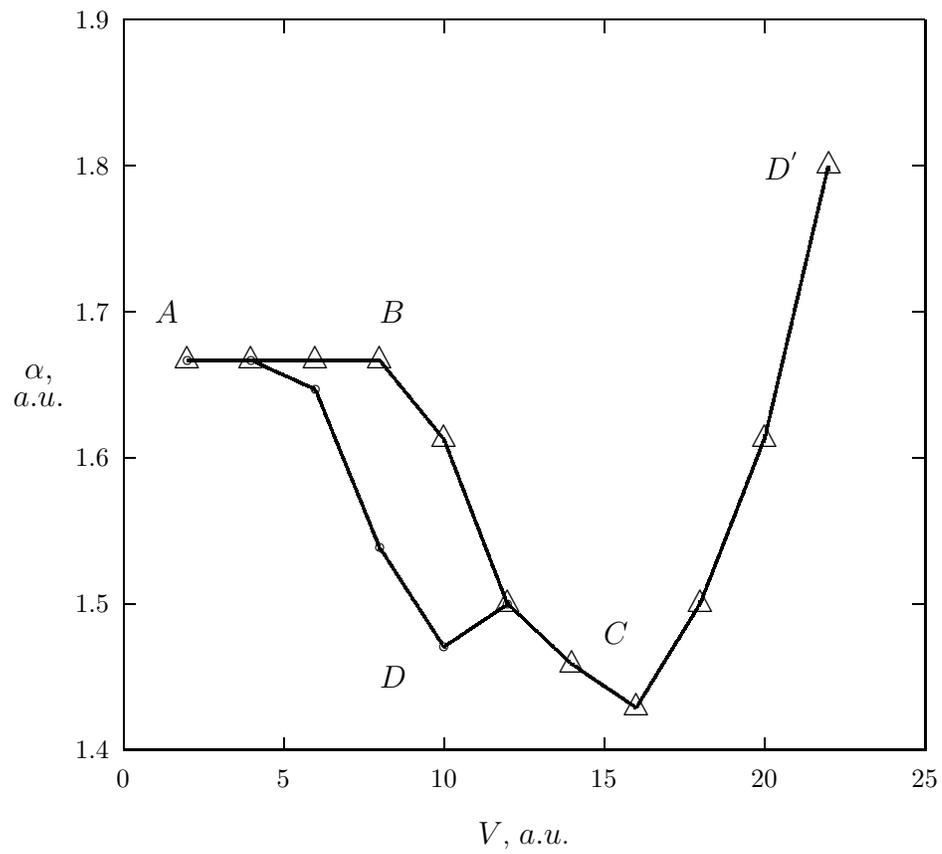
Figure~5: Experimental behaviour of US absorption by 2nd 
KCl sample. 

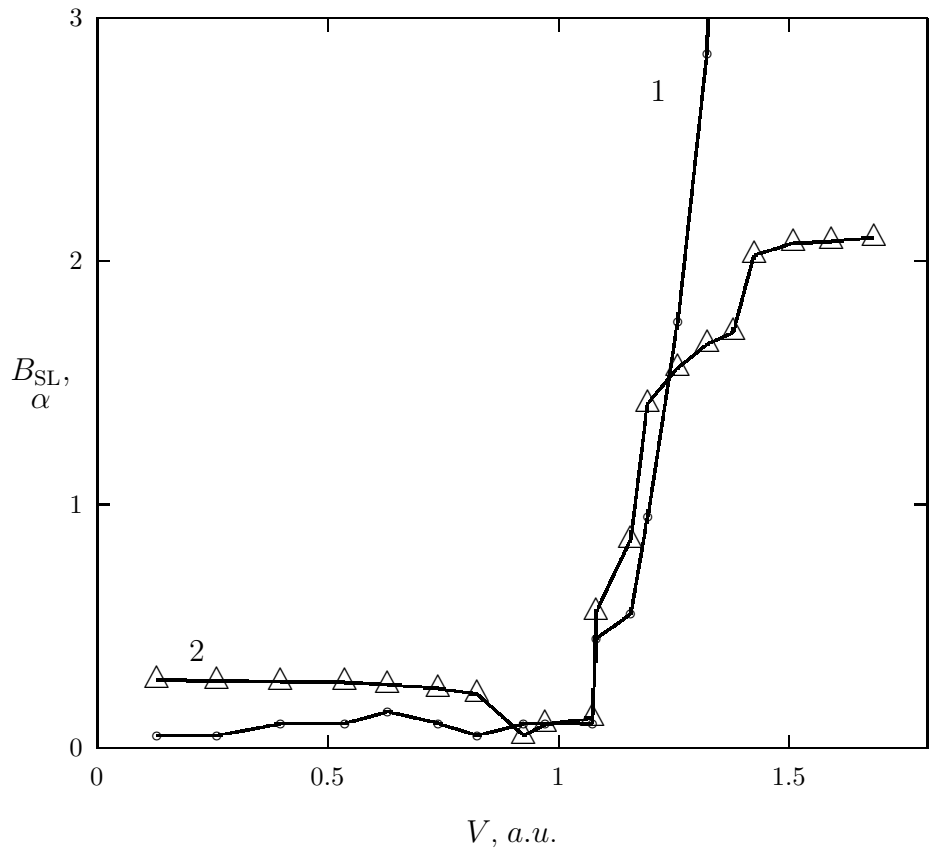
Figure~6: The comparison of threshold US amplitudes for SL
excitation and for continuous point defect generation. 

\clearpage

\begin{figure}[t]
\begin{center}
\setlength{\unitlength}{0.240900pt}
\ifx\plotpoint\undefined\newsavebox{\plotpoint}\fi
\sbox{\plotpoint}{\rule[-0.200pt]{0.400pt}{0.400pt}}%

\caption{}
\end{center}
\end{figure}

\clearpage

\begin{figure}[t]
\begin{center}
\setlength{\unitlength}{0.240900pt}
\ifx\plotpoint\undefined\newsavebox{\plotpoint}\fi
\sbox{\plotpoint}{\rule[-0.200pt]{0.400pt}{0.400pt}}%
%
\caption{}
\end{center}
\end{figure}

\clearpage

\begin{figure}[t]
\begin{center}
\setlength{\unitlength}{0.240900pt}
\ifx\plotpoint\undefined\newsavebox{\plotpoint}\fi
\sbox{\plotpoint}{\rule[-0.200pt]{0.400pt}{0.400pt}}%
%
\caption{}
\end{center}
\end{figure}

\clearpage

\begin{figure}[t]
\begin{center}
\setlength{\unitlength}{0.240900pt}
\ifx\plotpoint\undefined\newsavebox{\plotpoint}\fi
\sbox{\plotpoint}{\rule[-0.200pt]{0.400pt}{0.400pt}}%
%
\caption{}
\end{center}
\end{figure}

\end{document}